\documentstyle[12pt]{article}
\makeatletter
\def\eqalign#1{\null\vcenter{\def\\{\cr}\openup\jot\m@th
  \ialign{\strut$\displaystyle{##}$\hfil&$\displaystyle{{}##}$\hfil
      \crcr#1\crcr}}\,}
\makeatother

\begin{document}
\bigskip\bigskip\bigskip
\begin{center}
{\Large\bf 
On the discriminant of Harper's equation
}\\
\bigskip\bigskip\bigskip\bigskip
{\large I. V. Krasovsky}\\
\bigskip
Max-Planck-Institut f\"ur Physik komplexer Systeme\\
N\"othnitzer Str. 38, D-01187, Dresden, Germany\\
E-mail: ivk@mpipks-dresden.mpg.de\\
\medskip
and\\
\medskip
B.I.Verkin Institute for Low Temperature Physics and Engineering\\
47 Lenina Ave., Kharkov 310164, Ukraine.
\end{center}
\bigskip\bigskip\bigskip

\noindent
{\bf Abstract.}
The spectrum of Harper's equation is determined by the discriminant,
which is a certain polynomial of degree $Q$ if the commensurability parameter
of Harper's equation is $P/Q$, where $P$, $Q$ are coprime positive integers.  
A simple expression is indicated for the derivative of the discriminant
at zero energy for odd $Q$. Three dominant terms of the  
asymptotics of this derivative are calculated for the case of an arbitrary $P$
as $Q$ increases. The result gives a lower bound on the width
of the centermost band of Harper's equation and shows the effects
of band clustering.

It is noticed that the Hausdorff dimension of the spectrum is zero
for the case $P=1$, $Q$ infinitely large.

\newpage
\section{Introduction}
In the present paper we consider the following eigenvalue equation in
$l^2(Z)$ (see [\ref{Har}--\ref{AMS}] and [\ref{Si}--\ref{KS}] for the reviews):
\begin{equation}
\psi_{n-1}+2\cos(2\pi\alpha n +\theta)\psi_n
+\psi_{n+1}=\varepsilon\psi_n,\qquad
n=\dots,-1,0,1,\dots,\qquad \alpha,\theta\in{\bf R}\label{Harper}
\end{equation}
It is called Harper's equation with commensurability parameter $\alpha$. 
If $\alpha=P/Q$, where $P$ and $Q$ are relatively prime integers,
the spectrum of the corresponding operator consists of $Q$ intervals (bands).
The interesting case is the incommensurate one, i.e. when $\alpha$ is 
irrational. It is known that for irrational $\alpha$ (but not for rational)  
the spectrum is independent of $\theta$. There is a conjecture that for any 
irrational $\alpha$ the measure of the spectrum is zero. 

Let us call $\sigma(\alpha)$ the union
over all real $\theta$ of the spectra of (\ref{Harper}). 
Given (\ref{Harper}) with $\alpha=P/Q$, one can construct an independent of
$\theta$  polynomial $\Sigma(x)$ of degree $Q$ with the following property:
the spectrum  $\sigma(\alpha)$ is is the image of the interval 
$[-4,4]$ under the inverse of the transform $\Sigma(x)=\lambda$.
For convenience, we call
$\Sigma(x)$ the discriminant of Harper's equation.\footnote{
The usual definition of the discriminant of a periodic matrix corresponding to 
(\ref{Harper}) would be $S(x,\theta)=\Sigma(x)-2\cos\theta Q$.}
It is known that $\Sigma(x)=-\Sigma(-x)$ (hence $\Sigma(0)=0$)
for odd $Q$ and $\Sigma(x)=\Sigma(-x)$ for even $Q$. Also for even $Q$
we have $|\Sigma(0)|=4$ (which implies 
that $2$ bands of the spectrum merge in $0$). In the present paper, we will
give the expressions for the derivative at zero $\Sigma'(0)$ for odd $Q$.
Such expressions provide the lower bound on the width of the centermost band.
Using our approach, one can also calculate higher derivatives at zero,
but the formulas become more cumbersome.  

Several important results about Harper's equation with irrational $\alpha$
have been obtained using the properties of the discriminant.
Still not much is known about these properties. Here we briefly mention some 
results.
In 1992, Last and Wilkinson [\ref{LW}] noted the following fact about 
the derivatives of $\Sigma(x)$ at its zero crossings:
\begin{equation}
\sum_{k=1}^Q\frac{1}{|\Sigma'(x_k)|}={1\over Q},
\end{equation}
where $x_k$, $k=1,\dots,Q$ are the zeros of $\Sigma(x)$.
This property was subsequently used by Last [\ref{Last2}] to prove the 
vanishing of the measure of the spectrum for a certain class 
of irrational $\alpha$'s.
More detailed information on $\Sigma(x)$ can be obtained in the semiclassical 
case, that is when $\alpha$ is small [\ref{Wi},\ref{RB}] 
(also when $\alpha$ is close to a rational [\ref{Wi2}]). 
The uniform asymptotics is available for the case when
$P=1$ and $Q$ is large [\ref{Watson}]. From it, one can obtain the widths and
distribution of individual bands. The large $Q$ asymptotics can also be 
calculated for $P=2,3...$ [\ref{Watson}] but the expressions quickly become 
cumbersome 
with growing $P$. As we shall see, this is not the case with the asymptotics 
of $\Sigma'(0)$. We will calculate it for an arbitrary fixed $P$ up to the
terms decreasing with $Q$.
We will obtain several different types of asymptotic behaviour, which reflects
the effects of band clustering.

There is a considerable amount of literature on the question of what is
the Hausdorff dimension of $\sigma(\alpha)$ for irrational $\alpha$ 
([\ref{hausWi}] and references therein).
We make a note on this by showing that the Hausdorff dimension for the 
case of $P=1$ and infinite $Q$ is zero.

\section{General formula for the derivative of the discriminant at zero}
The main fact which will enable us to obtain an expression for $\Sigma'(0)$
is that the spectral problem (1) can be reduced to that for a matrix with 
zero main diagonal. In various avatars, this fact was noted in 
[\ref{KH0d},\ref{WZprl},\ref{WZmpl}].

Henceforth, we assume $\alpha=P/Q$, where $Q$ is odd, $P$ is relatively prime
with $Q$. It follows, e.g., from the reasoning in Appendix of [\ref{ba1}]
that $\Sigma(x)=-\det(L-xI)$, where $L$, $I$ are $Q\times Q$ matrices,
$I$ is the identity matrix, and the matrix elements of $L$ are the following:
$L_{k\;k-1}=L_{k-1\;k}=2\sin(\pi Pk/Q)$, if $k=1,2,\dots,Q-1$, and zero
otherwise. Thus, $L$ is a tridiagonal matrix with zero main diagonal.
It is therefore easy to get a simple expression for 
the derivative of the determinant
$\det(L-xI)$ at $x=0$. Using the identity 
$\prod_{k=1}^{(Q-1)/2}(2\sin(\pi P2k/Q))^2=Q$ at the last stage of the
calculation, we obtain
\begin{equation}
\Sigma'(0)=(-1)^{(Q-1)/2}Q\left(1+\sum_{k=1}^{(Q-1)/2}\prod_{j=1}^k
\frac{\sin^2(2j-1)\gamma}{\sin^2 2j\gamma}\right)
,\label{din0}
\end{equation}
where $\gamma=\pi P/Q$.
In the next section, we shall calculate the large $Q$ asymptotics of 
$\Sigma'(0)$ for the case of fixed $P$.

\section{Asymptotics for $\Sigma'(0)$}
Let $P$ be arbitrary fixed and consider the asymptotics of $\Sigma'(0)$
for large $Q$.
We can represent any odd $Q$ (relatively prime with $P$) in 
the form $Q=4Pr+s$, where $r,s$ are nonnegative integers, 
$s=1,3,5,\dots,4P-1$, excluding the numbers which are not relatively prime 
with $P$.   
Let
\begin{equation}
S=\sum_{k=1}^{(Q-1)/2}\prod_{j=1}^k
\frac{\sin^2(2j-1)\gamma}{\sin^2 2j\gamma}
\end{equation}
and rewrite it in the form
\begin{equation}
S=\sum_{m=1}^{(Q-s)/4P}\sum_{t=0}^{2P-1}
\prod_{j=1}^{t{Q-s\over 4P}+m}
\frac{\sin^2(2j-1)\gamma}{\sin^2 2j\gamma}+
\sum_{k=1}^{(s-1)/2}\prod_{j=1}^{{Q-s\over 2}+k}
\frac{\sin^2(2j-1)\gamma}{\sin^2 2j\gamma}.\label{s}
\end{equation}
To obtain the asymptotics, we first break the products here
into subproducts with the number of factors $(Q-s)/4P$. 
Then the sinuses become small for large $Q$ 
only in one connected part of each of the subproducts.
By a change of product variable $j=k(Q-s)/4P+j'$ for k even,
and $j=(k+1)(Q-s)/4P-j'+1$ for k odd, we obtain:
\begin{equation}
\prod_{j=1}^{t{Q-s\over 4P}+m}
\frac{\sin^2(2j-1)\gamma}{\sin^2 2j\gamma}=
A_0 A_1\cdots A_{t-1} 
\prod_{j=1}^{m}
\frac{\sin^2(2j-1+\delta_1)\gamma}
{\sin^2(2j+\delta_2)\gamma}
\end{equation}
where $\delta_1=\delta_2=-ts/2P$ for t even;  $\delta_1=(t+1)s/2P$,
$\delta_2=-1+(t+1)s/2P$ for t odd; and 
\begin{equation}
\eqalign{
A_{2k}=\prod_{j=1}^{(Q-s)/4P}
\frac{\sin^2(2j-1-ks/P)\gamma}
{\sin^2(2j-ks/P)\gamma},\\
A_{2k-1}=\prod_{j=1}^{(Q-s)/4P}
\frac{\sin^2(2j-1+ks/P)\gamma}
{\sin^2(2j-2+ks/P)\gamma}.}
\end{equation}
Substituting these expressions into (\ref{s}) and making a change of variables
$m=(Q-s)/4P-m'$ in one of the sums, we finally obtain the following expression
that will be a starting point for deriving the asymptotics:
\begin{equation}
\eqalign{
S=\sum_{t=0}^{P-1}\prod_{i=0}^{2t-1}A_i\left\{
\sum_{m=1}^{(Q-s)/4P}\left(\prod_{j=1}^m
\frac{\sin^2(2j-1-ts/P)\gamma}
{\sin^2(2j-ts/P)\gamma}+\right.\right.
\\
\left.\left.
A_{2t}A_{2t+1}\prod_{j=1}^m
\frac{\sin^2(2j-2+(t+1)s/P)\gamma}
{\sin^2(2j-1+(t+1)s/P)\gamma}
\right)+ A_{2t}(A_{2t+1}-1)\right\}+\\
\prod_{i=0}^{2P-1}A_i
\sum_{m=1}^{(s-1)/2}\prod_{j=1}^m
\frac{\sin^2(2j-1-s)\gamma}
{\sin^2(2j-s)\gamma}.}\label{s2}
\end{equation}
Henceforth, we assume $\prod_{i=1}^0f_i\equiv 1$,
$\sum_{m=1}^0g_m\equiv 0$.

We will need to estimate asymptotics of 2 types of products.
Let
$c_j=\sin^2(2j+2a)\gamma/\sin^2(2j+2b)\gamma$, where $a$ and $b$ are any 
constants such that $c_j$ ($j=1,2,\dots$) are finite. 
First, for $k(Q)$ such that $k/Q\to 0$ as $Q\to\infty$, we have:
\begin{equation}
\prod_{j=1}^k c_j=
\frac{\Gamma^2(b+1)}{\Gamma^2(a+1)}
\frac{\Gamma^2(k+a+1)}{\Gamma^2(k+b+1)}(1+O(Q^{-2+\epsilon})),\label{a1}
\end{equation}
where we can choose $\epsilon$ arbitrary small if $k/Q\to 0$ fast enough.
Secondly, if $M\to\infty$, but $M/Q\to 0$ as $Q\to\infty$, we have
\begin{equation}
\eqalign{
\prod_{j=M+1}^k c_j=\\
\exp\left\{2\sum_{j=M+1}^k\ln\frac{1-(2a\gamma)^2+O(\gamma^4)+
(2a\gamma+O(\gamma^3))\cot 2j\gamma}
{1-(2b\gamma)^2+O(\gamma^4)+
(2b\gamma+O(\gamma^3))\cot 2j\gamma}\right\}=\\
\left(\frac{\sin(2k+1)\gamma}{2\gamma M}\right)^{2(a-b)}
\left(1+2\gamma\left(a^2-b^2\right)[\cot(2k+1)\gamma-
\cot(2M+1)\gamma]+\right.\\
\left. O(1/Q^2)+O(1/MQ)\right).}\label{a2}
\end{equation}
Here we expanded the logarithm in series using the fact that $1/M$ is a small
parameter, and then applied the Poisson summation formula to estimate the sum.

Now let us estimate the asymptotics of $A_k$:
\begin{equation}
\eqalign{
A_{2k}=\prod_{j=1}^M
\frac{\sin^2(2j-1-ks/P)\gamma}
{\sin^2(2j-ks/P)\gamma}
\prod_{j=M+1}^{(Q-s)/4P}
\frac{\sin^2(2j-1-ks/P)\gamma}
{\sin^2(2j-ks/P)\gamma}=\\
2\gamma\frac{\Gamma^2(1-ks/2P)}{\Gamma^2(1/2-ks/2P)}(1+O(1/Q^2)),}\label{Aae}
\end{equation}
where we used (\ref{a2}) and (\ref{a1}) for $b=-ks/2P$, $a=b-1/2$
(in (\ref{a1}) we took the asymptotics of $\Gamma(x)$ for large $x$).

Similarly, using (\ref{a1}) and (\ref{a2}) for $b=-1+ks/2P$, $a=b+1/2$,
we get:
\begin{equation}
A_{2k-1}=\frac{1}{2\gamma}
\frac{\Gamma^2(ks/2P)}{\Gamma^2(1/2+ks/2P)}(1+O(1/Q^2)).\label{Aao}
\end{equation}

The last nontrivial step in our calculations is the following estimate
for $a=b-1/2$:
\begin{equation}
\eqalign{
\sum_{k=1}^{(Q-s)/4P}\prod_{j=1}^k c_j=
\sum_{k=1}^M\prod_{j=1}^k c_j+
\sum_{k=M+1}^{(Q-s)/4P}\prod_{j=1}^M c_j\prod_{j=M+1}^k c_j=\\
\frac{\Gamma^2(b+1)}{\Gamma^2(b+1/2)}\left\{
\eta(b)-\ln\gamma+(4P-s){\pi\over 2Q}+\gamma\left(2b-{1\over 2}\right)
+O(1/Q^2)\right\},}\label{sum-est}
\end{equation}
where the constant of Euler's type
\begin{equation}
\eta(b)=\lim_{M\to\infty}\left(
\sum_{k=1}^M\frac{\Gamma^2(k+b+1/2)}{\Gamma^2(k+b+1)}
-\ln M\right).\label{const}
\end{equation}
Such constants were studied in [\ref{ES}].\footnote{I am grateful to R. Askey
for indicating this reference.}

To obtain (\ref{sum-est}) we used expressions (\ref{a1},\ref{a2}),
and then applied the Poisson summation formula to get the asymptotics of 
the sum. The most important moment in our derivation is to choose $M$ in such
a way that $M\to\infty$ and $M/Q\to 0$ as $Q\to\infty$. Apart from this
condition, $M$ is arbitrary. Naturally, $M$ cancels in the final expression.

Using (\ref{sum-est},\ref{Aae},\ref{Aao}) and (\ref{a1}),  
it is easy to get the asymptotic formula for $S$ from (\ref{s2}).
As a result, we find that the derivative of the discriminant 
$\Sigma'(0)=(-1)^{(Q-1)/2}Q(1+S)$ has the following asymptotics
as $Q\to\infty$:
\begin{equation}
\eqalign{
\Sigma'(0)=(-1)^{(Q-1)/2}Q\left(
{1\over\pi}\sum_{t=0}^{P-1}\prod_{i=1}^t\cot^2\left[\frac{\pi si}{2P}
\right]
\left\{2\ln\frac{Q}{\pi P}+\eta\left(-{st\over 2P}\right)+
\right.\right.
\\
\left.
\eta\left({s-1\over 2}-{st\over 2P}\right)+
\frac{\Gamma^2(s/2-st/(2P))}{\Gamma^2((s+1)/2-st/(2P))}\right\}+\\
\left.
1+{1\over\pi}\sum_{m=1}^{(s-1)/2}\frac{\Gamma^2(m-1/2)}{\Gamma^2(m)}
\right)+ o(1)}\label{A}
\end{equation}
where the constants $\eta(b)$ are defined in (\ref{const}).

It is known that the bands for large $Q$ appear away from zero energy in 
clusters of $P$ bands separated by wide gaps. In a cluster around zero 
energy there are $s$ bands. We see from (\ref{A}) that at least to the 
leading $Q\ln Q$ order there are 
no more than $P/2$ types of different asymptotic behaviour (because of the
periodicity of cotangent). Namely, for odd $P>1$, $s=1,3,\dots,P-2$;
for even $P$,  $s=1,3,\dots,P-1$ excluding the values not relatively prime
with $P$. It would be interesting to prove if this is the case for the whole
asymptotic series.

For $P=1$ there is only one type of the asymptotic behaviour:
\begin{equation}
\Sigma'(0)={2\over\pi}(-1)^{(Q-1)/2}Q\left(
\ln{Q\over\pi}+\eta(0)+\pi\right)+o(1).\label{dp=1}
\end{equation}

\section{Hausdorff dimension}
Consider the case $P=1$, $Q\to\infty$. We are interested in the Hausdorff
dimension of the limiting spectrum (there is no reason for it to be equal to
the dimension for the case $P/Q=0$, which is, obviously, $1$). 
The spectrum of Harper's equation lies within the interval $[-4,4]$.
The dominant term of the uniform asymptotics for the discriminant 
$\Sigma(x)$ on $x\in(-4,4)$ as $Q\to\infty$ can be found in [\ref{Watson}]. 
(One can also give a simpler derivation of (\ref{d1}) taking the representation
of $\Sigma(x)$ in terms of $\det(L-xI)$ and matching the semiclassical
solution with the asymptotics for Meixner-Pollaczek polynomials [\ref{unpub}].)

For $|x|\ll 1$, ($\lambda=x/4$)
\begin{equation}
\Sigma(x)\sim
4\cosh(\lambda Q)\cos\left(
\frac{2\lambda Q}{\pi}\ln{4Q\over\pi}-2\arg\Gamma\left({1\over 2}+i
{\lambda Q\over\pi}\right)-{\pi Q\over 2}\right)\label{d1}
\end{equation}
For $xQ\gg 1$, $x\le 4-\epsilon<4$ (for any $\epsilon>0$) 
\begin{equation}
\Sigma(x)\sim 2e^{2Q\mu}\cos(2Q\nu),\label{d2}
\end{equation}
where
\begin{equation}
\eqalign{
\mu=
\int_{(2\pi)^{-1}\arccos(2\lambda-1)}^{1/2}
{\rm arccosh}(2\lambda-\cos 2\pi t)dt,\\
\nu=
\int_0^{(2\pi)^{-1}\arccos(2\lambda-1)}
\arccos(2\lambda-\cos 2\pi t)dt,}
\end{equation}
the branch of arccosine being $0\le\arccos x\le\pi$ for $x\in[-1,1]$.
Given (\ref{d2}), the asymptotics for the symmetric region of negative $x$
is obvious because of the property $\Sigma(x)=(-1)^Q\Sigma(-x)$.

It is interesting to compare $\Sigma'(0)$ obtained from (\ref{d1}) with
(\ref{dp=1}). This gives the representation $\eta(0)=\ln 16+C-\pi$, where $C$
is Euler's constant.

In the region where $|xQ|\gg 1$, the bands of the spectrum are exponentially
narrow in $Q$, and their density, as one can deduce, is given by
\begin{equation}
\rho(x)=K'(x/4)/(2\pi^2),\label{rho} 
\end{equation}
where $K'(\lambda)=K(\sqrt{1-\lambda^2})=
\int_0^{\pi/2}(1-(1-\lambda^2)\sin^2u)^{-1/2}du$ is the 
complete elliptic integral. Indeed, we see from (\ref{d2}) that $\rho(x)$
is given by the expression $\rho(x)=(1/\pi)2 |d\nu/d\lambda| d\lambda/dx$.
Further, we have 
\begin{equation}
\left|{d\nu\over d\lambda}\right|=
{1\over\pi}\int_0^{\arccos(2\lambda-1)}\frac{dt}
{\sqrt{1-(2\lambda-\cos t)^2}}.
\end{equation}
Making the change of variable $\cos t=(1-\lambda)u+\lambda$, we get
\begin{equation}
\left|{d\nu\over d\lambda}\right|=\frac{2/\pi}
{1+\lambda}K\left(\frac{1-\lambda}{1+\lambda}\right)=
\frac{1}{\pi}K'(\lambda),
\end{equation}
which yields (\ref{rho}). 
(One can also deduce (\ref{rho}) from the results of [\ref{ba1}]).
This expression for $\rho(x)$ as $\alpha=P/Q\to 0$
is very natural as it is exactly the well-known density of states for the
case $\alpha=0$.

Because of the exponentially small widths of the bands, it is easy to show
that the Hausdorff dimension for the region $|xQ|\gg 1$ of the spectrum 
is equal to zero. (Note that for $x\in[4-\epsilon,4]$, $0<\epsilon\ll 1$,
the bands remain exponentially small. The expressions for the positions
of these bands to several orders in $1/Q$ are given in [\ref{RB}].) 
In the other region, $|x|\ll 1$, the widths of the bands are of order
$1/Q\ln Q$ when $|xQ|\sim 1$ as we see from (\ref{d1}). 
More precisely, the dominant asymptotic
term $\delta(\lambda Q)$ for the width of the bands in the neighbourhood of 
$\lambda Q$ is
\begin{equation}
\delta(\lambda Q)={4\pi\over Q\ln Q}
\arcsin{1\over\cosh(\lambda Q)}.\label{width}
\end{equation}
The factor 4 appears here because we are looking for the width ``on the scale''
$x$, while $\lambda=x/4$.
  
If we sum the widths $\delta_i$ of all these bands, we will obtain the famous
Thouless formula [\ref{T},\ref{Watson}] for the total bandwidth $W$. Indeed,
if $t=\lambda Q$, then in the interval $dt$ there are, to the main order,
${2\over\pi}\ln Q{dt\over\pi}$ bands. Hence, 
\begin{equation}
W=\sum\delta_i\sim{16\over\pi Q}\int_0^\infty\arcsin{1\over\cosh t}dt=
{32\over\pi Q}\beta(2),
\end{equation}
where $\beta(2)$ is Catalan's constant.
To estimate the Hausdorff dimension, we consider the sum
$W(d)=\sum\delta_i^d$. We have
\begin{equation}
W(d)\sim{1\over Q^d\ln^{d-1}Q}{4^{1+d}\over\pi^{2-d}}
\int_0^\infty\arcsin^d{1\over\cosh t}dt,
\end{equation}
which tends to zero as $Q\to\infty$ for any $d>0$ and is infinity for $d=0$.
This implies that the Hausdorff dimension is zero.
Thus, the Hausdorff dimension of the whole spectrum for $P=1$, $Q\to\infty$ is 
zero. 
 
\section{Acknowledgements}
I am grateful to R. Askey, J. Bellissard, P. Wiegmann, 
and M. Wilkinson for useful discussions.

\end{document}